# Microwave Chirality Imaging for the Early Diagnosis of Neurological Degenerative Diseases

Wending Mai, *Senior Member, IEEE,* and Yifan Chen, *Senior Member, IEEE*

*Abstract*— We propose a system to visualize the chirality of the protein in brains, which would be helpful to diagnose early neurological degenerative diseases in *vivo*. These neurological degenerative diseases often occur along with some mark proteins. By nanoparticle instilling and metamaterial technique, the chiral effect of the mark proteins is assumed to be manifest in microwave regime. Therefore, by detecting the transmission of cross-polarization, we could detect the chirality that rotates the microwave polarization angle. We developed a numerical method to simulate the electromagnetic response upon chiral (bi-isotropic) material. Then a numerical experiment was conduct with a numerical head phantom. A map of cross-polarized transmission magnitude can be reached by sweeping the antenna pair. The imaging results matches well with the distribution of chiral materials. It suggests that the proposed method would be capable of *in vivo* imaging of neurological degenerative disease using microwaves.

## I. Introduction

NEUROLOGICAL degenerative diseases are getting increasing research interest as the average length of life increases. The two most important neurological diseases are the Alzheimer's disease (AD) and the Parkinson's disease (PD). Neurological degenerative diseases are the result of a combination of genetic, lifestyle, and environmental factors, caused in part by specific genetic changes. Eventually, the nerve cells in the patient's brain shrink or even die. The transformation of the nerve cells at their early stage is so subtle that undetectable by modern radiology techniques (CT, MRI for example). Currently, the diagnosis of neurological degenerative diseases in early stage is still mainly based on the patients' behavior, which is very inaccurate.

It has been found that the typical histopathological changes of AD are amyloid deposits and neuronal protein tangles in the brain, and there are currently multiple theories being attempted to explain this change, including the β-amyloid (Aβ) waterfall theory, tau protein theory, and neurovascular hypothesis [1-4]. Similarly, PD are found dependence with the existence of α-synuclein amyloidogenic fiber. If by somehow the existence of these mark fibers can be identified, it would help the earlier diagnose of neurological degenerative diseases.

Scientists discovered that some substances were able to rotate the polarization of linearly-polarized light. This phenomenon is known as optical activity, or polarization rotation, which has been widely applied to the measurement of chiral molecule concentrations in compounds and solutions [5-7]. The polarization rotation effect depends both on the chirality and concentrations of the chiral molecules. This kind of polarization detection technique is widely used in optics to determine the concentrations of simple sugars, glucose, fructose, etc.

Studies have also shown that the existence of some mark proteins will bring about changes in chirality in optical regime. Kumar et al. developed a chirality detection method for α-synuclein protein. With golden nanoparticles, the protein shows surface plasmonic phenomenon. The chirality from the helical structure of protein brings optical activity. They further showed that this technique can be applied to the detection of PD where α-synuclein amyloidogenic fiber dominates [1]. Other research shows that the dominating protein of AD has a similar phenomenon [8-10]. These existing works show that detecting the chirality of specific types of proteins can help to diagnose the neurological degenerative diseases. However, those studies are carried out in the optical regime. Light cannot transmit through the skull. Therefore, in order to achieve in *vivo* detection, skull optical clearing window is required, which however is not suitable for early detection [11].

On the other hand, microwave can transmit through the skull and other human tissues. Based on that, many microwave-based medical imaging techniques have been developed to detect breast cancer, stroke, etc. It has been found that the optical activity of chiral proteins is mainly in optical regime. But by using twisted optical metamaterials and deference method, the optical activity of chiral proteins can be observed in much lower frequencies. Patterson et al. further lowered the frequency by detecting molecular chirality using nonlinear microwave spectroscopy [12].

In this work, we visualize the distribution of the chiral mark proteins to help diagnosis of neurological degenerative disease. We propose a system with a pair transmitting and receiving antennas with polarization direction orthogonal with each other. The transmission of cross-polarization is measured which represents the chirality and concentrations of chiral proteins in the human brain. By sweeping the pair of antennas, we can get a distribution map of chiral mark proteins. Then the chirality distribution is visualized. The proposed imaging technique can be helpful for the early diagnosis of neurological degenerative diseases in *vivo*.

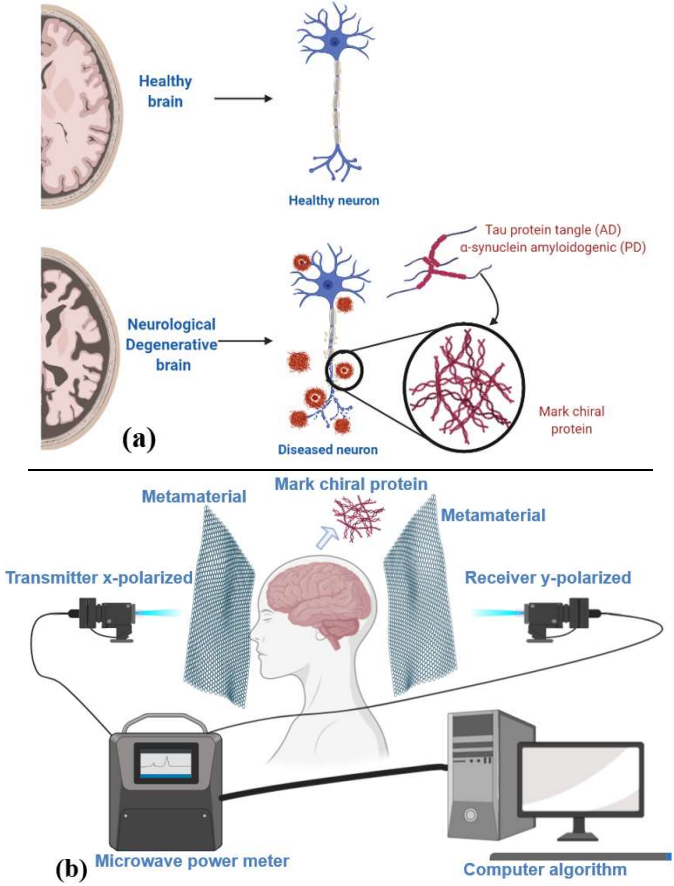

Fig. 1. Illustration of (a) pathology of neurological degenerative diseases, and (b) proposed in *vivo* detection system in microwave.

## II. METHODOLOGY

Studies show the neurological degenerative diseases are often related with some mark proteins. As shown in Fig. 1 (a), the tau and α-synuclein amyloidogenic proteins are tangled surrounding the diseased neurons and decompose their functionality. The mark protein is in larger scale and stronger chirality than healthy proteins. Therefore, by detecting the chirality of the human brain, the existence of those mark proteins can be identified.

To diagnose the neurological degenerative disease in vivo, we propose a microwave detection system depicted in Fig. 1 (b). A pair of antennas were placed at the opposite sides of the patient's head with orthogonal polarization direction. Twisted optical or nonlinear metamaterials are placed between the antennas and the human head to increase the chiral sensibility of molecules in microwave regime. The receiver antenna is connected to a microwave power meter to detect how much power was transmitted in cross-polarization.

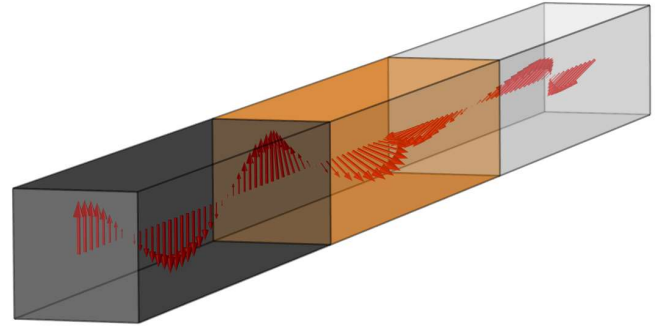

Fig. 2. Illustration of a segment of chiral material (orange) that rotates the polarization of the incident wave.

### A. Governing equation of chiral material

Considering Maxwell's symmetric curl equations for linear isotropic homogeneous medium:

$$j\omega \vec{B} = -\nabla \times \vec{E} \quad (1)$$
$$j\omega \vec{D} = \nabla \times \vec{H} \quad (2)$$

The general chiral non-reciprocal medium is described by the constitutive relations:

$$\vec{D} = \varepsilon \vec{E} + \xi \vec{H} \quad (3)$$
$$\vec{B} = \zeta \vec{E} + \mu \vec{H} \quad (4)$$

where,

$$\xi = (\chi - j\kappa)\sqrt{\varepsilon_0 \mu_0} \quad (5)$$
$$\zeta = (\chi + j\kappa)\sqrt{\varepsilon_0 \mu_0} \quad (6)$$

The chirality parameter κ is a dimensionless quantity which defines the degree of handedness of the material. In a racemic mixture where the chiral materials of both types *i.e.* right and left handed are equal and for nonchiral materials the chirality parameter κ goes to zero. The Tellegen parameter χ describes the magnetoelectric effect. A nonzero value of χ means that the material is non-reciprocal.

To solve the problem using finite element method (FEM), the constitutive relations need to be integrated with electric Helmholtz equation. By conducting curl operation on Eq. (1), we have:

$$\nabla \times \nabla \times \vec{E} + j\omega \nabla \times \vec{B} = 0 \quad (7)$$

By substituting Eq. (5) with Eq. (4), we have:

$$\nabla^2 \vec{E} + j\omega \zeta \nabla \times \vec{E} + \omega^2 \mu \vec{D} = 0 \quad (8)$$

It can further go as:

$$\nabla^2 \vec{E} + 2\tilde{a}(\nabla \times \vec{E}) + \tilde{k}_0^2 \vec{E} = 0 \quad (9)$$

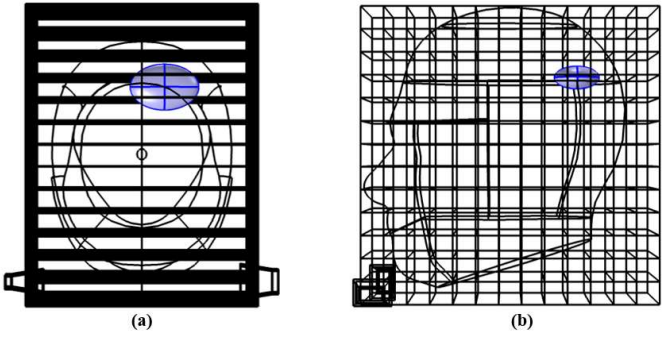

Fig. 3. Illustration of the imaging system in (a) top view and (b) front view.

where

$$\tilde{a} = \omega\kappa\sqrt{\varepsilon_0\mu_0} \tag{10}$$

$$\tilde{k}_0 = \omega\sqrt{\varepsilon\mu - \varepsilon_0\mu_0(\chi^2 + \kappa^2)} \tag{11}$$

If the medium is nonchiral and reciprocal, we have $\tilde{a} = 0$ and $\tilde{k}_0 = k_0 = \omega\sqrt{\varepsilon\mu}$, which brings Eq. (9) as the conventional electric Helmholtz equation. As can be seen, our derivation is general for both/either chiral and nonreciprocal medium. But in this paper, we only investigate the chiral and reciprocal medium by setting $\chi = 0$.

Above equations can be incorporated into commercial FEM software COMSOL by modifying the governing equation of chiral medium. As mentioned, the derivation is based on linear isotropic homogeneous chiral materials. COMSOL employs the finite element method (FEM) for numerical simulation. It possesses great flexibility by allowing users to edit the equations.

To validate the derived governing equation of chiral materials, we firstly investigate a vertically polarized plane wave propagating in an air box. As shown in Fig. 2, a segment of chiral medium (orange) is placed in the middle of the air box. The chiral material has $\varepsilon_r = \mu_r = 1$, and the chiral parameter $\kappa = 0.5$. The red arrows represent the electric field. It shows that the wave polarization is rotated from the input port (black) to the output port (white). This is within expectation as the chirality rotates the polarization and hence increases the transmission of the cross-polarization.

*B. The visualization of chirality distribution*

To investigate the chirality of protein in microwave regime, we assume that several approaches have been applied to strengthen the chiral effect. These include nano-particles instilling, twisted optical metamaterials or nonlinear metamaterials placed before the antennas, etc.

Fig. 3 (a) shows the simulation setup. The human head is submerged in a box filled with a matching medium for better transmission. The matching medium is designed as lossless, with $\varepsilon_r = 53, \mu_r = 1$, which is the average value of human brain at $f_0 = 2.45 MHz$.

The numerical head phantom uses the geometry of the SAM Phantom provided IEEE, with homogenized with $\varepsilon_r = $

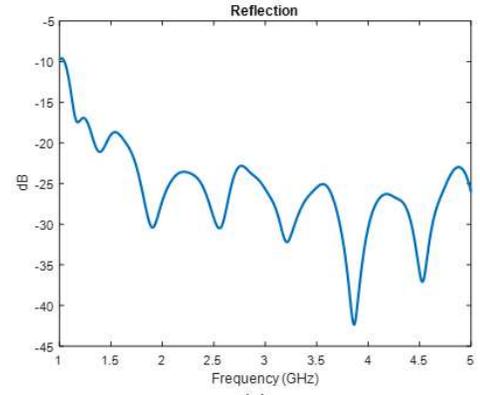

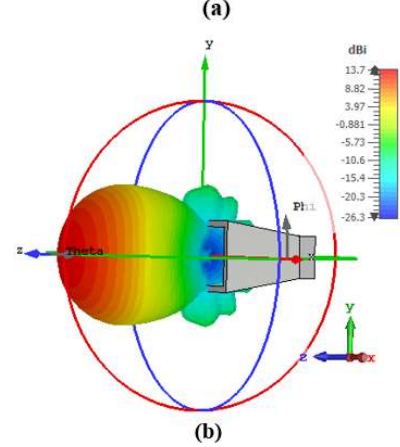

Fig. 4. (a) The reflection and (b) far-field directionality of the horn antenna in the matching medium.

$53, \mu_r = 1, \sigma = 1.1\ S/m$. Inside the brain, we placed an ellipsoid representing the neurology degenerative brain filled with chiral proteins. The ellipsoid has semiaxis of 30, 20 and 10 mm, respectively. Its chiral parameter is set as: $\kappa = 0.5$.

To visualize the chirality distribution inside of human head, we can plot the map of cross-polarization in a grid manner. A pair of antennas are placed in opposite direction as shown in Fig. 3. The antennas are designed as rectangular horn antennas with polarization direction orthogonal with each other. If the material between those two antennas is non-chiral, the transmission of the cross-polarization should be almost zero. But if the material is chiral, the polarization of the incident wave will be rotated, and we could expect that the transmission of cross-polarization will increase accordingly.

Besides linear polarization, rectangular horn antenna has high directionality. Therefore, we can assume that most of the electromagnetic energy are concentrated in the cuboidal tube between two antennas. Consequently, the cross-polarized transmission represents the material chirality inside. Fig. 3 (a) and (b) depicts the head phantom, the antennas and the visual cells in top view and front view, respectively. By sweeping the antennas long each cell, we can get a map of cross-polarized transmission. As shown in Fig. 4 (b), the grid has $12 \times 12$ cells, with resolution of 20 mm.

The antennas are submerged in the matching medium. The input side is a WR42 rectangular waveguide $10.668 \times 4.318\ mm$. The output side has dimension of $14.752 \times$ 9.752 mm. The length of the horn is 36 mm. Fig. 4 (a) and (b) depicts the reflection and far-field directionality of the horn antenna in the matching medium. The antenna has directionality of 13.7 dBi at 2.45 GHz.

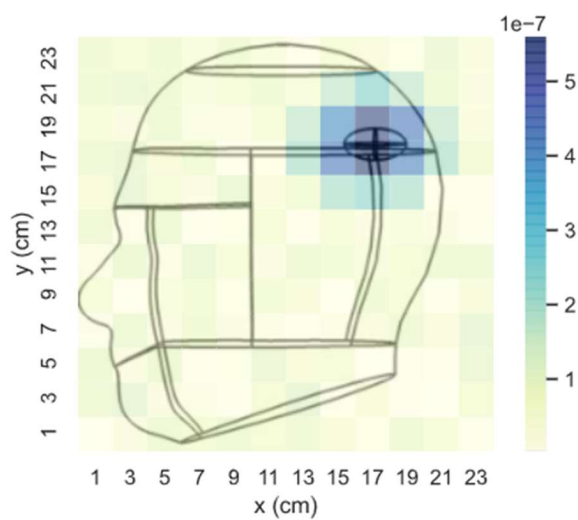

Fig. 5. The simulated map of cross-polarized transmission, which indicates the distribution of chiral mark proteins.

By sweeping the pair of antennas, we can investigate the grid distribution of cross-polarized transmission as shown in Fig. 5. The *x*- and *y*- label indicate the center of grid cell. The map of the cross-polarized transmission magnitude matches well with the distribution of chiral materials. Besides, good contrast is observed between chiral and non-chiral materials. Therefore, it is promising that by detecting the transmission magnitude of cross-polarization, the chiral mark proteins can be detected, which helps the early diagnosis of neurological degenerative diseases.

### III. DISCUSSION

In this paper, the microwave imaging technique is utilized to visualize the distribution of chiral mark proteins that helpful for the early diagnosis of neurological degenerative diseases. The result shows that microwave chirality imaging is a promising technique to detect and diagnose the neurological degenerative disease at the early stage.